\documentclass[twocolumn,showpacs,preprintnumbers,amsmath,amssymb]{revtex4}

\usepackage{graphicx}
\usepackage{dcolumn}
\usepackage{bm}

\begin{document}

\title{Scaling behavior of an artificial traffic model on scale-free networks}
\author{Shi-Min Cai$^{1}$}
\author{Gang Yan$^{1}$}
\author{Tao Zhou$^{1,2}$}
\email{zhutou@ustc.edu}
\author{Pei-Ling Zhou$^{1}$}
\author{Zhong-Qian Fu$^{1}$}
\author{Bing-Hong Wang$^{2}$}

\affiliation{$^{1}$Department of Electronic Science and Technology,
$^{2}$Department of Modern Physics, \\ University of Science and
Technology of China, Hefei Anhui, 230026, P. R. China}

\date{\today}

\begin{abstract}
In this article, we investigate an artificial traffic model on
scale-free networks. Instead of using the routing strategy of the
shortest path, a generalized routing algorithm is introduced to
improve the transportation throughput, which is measured by the
value of the critical point disjoining the free-flow phase and the
congested phase. By using the detrended fluctuation analysis, we
found that the traffic rate fluctuation near the critical point
exhibits the $1/f$-type scaling in the power spectrum. The
simulation results agree very well with the empirical data, thus
the present model may contribute to the understanding of the
underlying mechanism of network traffics.
\end{abstract}


\maketitle

\section{Introduction}
Complex networks can be used to describe a wide range of systems
from nature to society, thus there has been a quickly growing
interest in this area since the discoveries on small-world phenomena
\cite{Watts} and scale-free properties \cite{Barabasi}. Due to the
increasing importance of large communication networks upon which our
society survives, the dynamical processes taking place upon the
underlying structures such as traffics of information flow have draw
more and more attentions from the physical and engineering
communities. Previous studies on understanding and controlling
traffic congestion on network have a basic assumption that the
network has a homogeneous structure \cite{Li,Leland,Taqqu,Crovella}.
However, many real communication networks, including the Internet
\cite{Pastor} and WWW \cite{Albert1999}, display the scale-free
property. Therefore it is of great interest to explore how this
highly heterogenous topology affects the traffic dynamics, which
have brought a number of research works in recent years (see the
review paper \cite{Tadic2006} and the references therein).

To improve the transportation efficiency on complex networks,
Guimer\'{a} \emph{et al.} presented a formalism that can deal
simultaneously with the searching and traffic dynamics in parallel
transportation systems \cite{Guimera}. This formalism can be used
to optimize network structure under a local search algorithm
through the knowledge of the global information of whole network.
By using a global and dynamical searching algorithm aimed at the
shortest paths, Zhao \emph{et al.} provided the theoretical
estimates of the communication capacity \cite{Zhao2005}. Holme
proposed a much intelligent routing protocol in which the packet
can detour at obstacle thus guarantee much better performance than
the traditional routing strategy which just waits at obstacle
\cite{Holme2003}. Since the global information is usually
unavailable in large-scale networks, Tadi\'{c} \emph{et al.}
proposed a traffic model on a mimic WWW network \cite{Tadic2001}
wherein only local information, is available. This work
highlighted the relationship of global statistical properties and
microscopic density fluctuations \cite{Tadic2002,Tadic2004}. In
addition, Yin \emph{et al.} \cite{Yin2006,Wang2006} investigated
how the local routing protocol affects the traffic condition. The
routing strategies for the Internet \cite{Echenique} and
disordered networks \cite{Braunstein} are also studied.

\begin{figure}
\begin{center}
\scalebox{1}[0.8]{\includegraphics{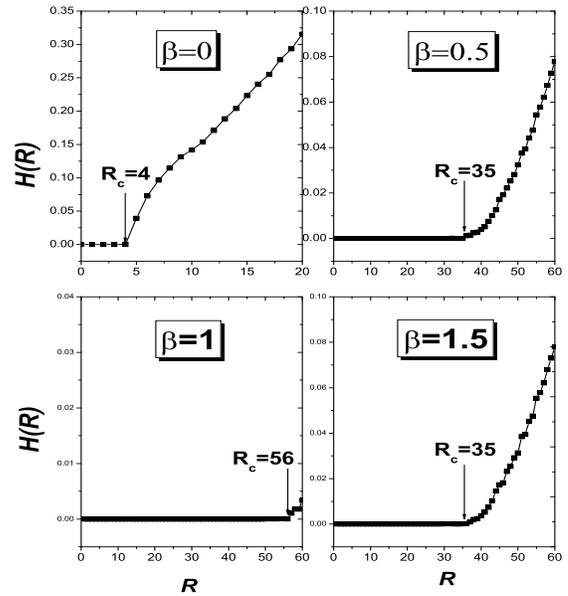}} \caption{The order
parameter $H$ vs $R$ for BA networks with average degree $\langle
k \rangle =4 $ and of size $N=1500$. The estimative values of
critical point $R_{c}$ of phase transition are 10, 35, 56 and 35
for $\beta=$0.3, 0.5, 1 and 1.5, respectively. }
\end{center}
\end{figure}

Recent empirical studies on communication networks have found
pervasive evidence of some surprising scaling properties. One
example of such discoveries is that the traffic rates of a given
link in the Internet (i.e. the number of packets or bytes
traversing a given link per time unit) exhibit the self-similar
(or fractal-like) scaling, and the multifractal scaling is also
found over small time scales
\cite{Crovella,Leland1994,Paxson1997,Feldmann,Yang2006}. These
empirical studies describe pertinent statistical characteristics
of temporal dynamics of measured traffic rate process and provide
ample evidence that these traces are consistent with long-range
correlated behavior. Faloutsos \emph{et al.} concerned the
autonomous system (AS) of the Internet and pointed out for the
first time that the degree distribution of snapshots in the AS
level follows a heavy-tailed function that can be reasonably
approximated by power-law forms \cite{Faloutsos}. Following these
pioneering works, Percacci \emph{et al.} found that the round trip
time (RTT) or Internet delay time can also be characterized by a
slow power-law decay \cite{Percacci}. This finding implies that
network inter-connectivity topology structure of Internet and
transportation delay both have a highly heterogenous
characteristic. Thus, the assumption of Poisson process, which had
been a major traffic model with homogeneous network in the
traditional traffic theory, has clearly lost its validity. This is
also the very reason why in this article we use scale-free
networks for simulations. Furthermore, the observation of a phase
transition between the free-flow phase and the congested phase in
the Internet traffic is demonstrated by Takayasu \emph{et al.}
through both the RTT experiment \cite{Takayasu1996,Fukuda1999} and
packet flow fluctuation analysis \cite{Takayasu1999,Takayasu2000}.
However, the $1/f$-type scaling is only detected near the critical
state \cite{Takayasu1996,Fukuda1999,Takayasu1999,Takayasu2000}.

In this article, we investigate a traffic model on scale-free
networks, in which packets are routed according to the global
topological information with a single tunable parameter $\beta$.
Instead of using the routing strategy based on the shortest paths,
we give a generalized routing algorithm to find the so-called
\emph{efficient path} to improve the transportation efficiency
that is measured by the value of the critical point disjoining the
free-flow and the congested phases. Furthermore, by using the
detrended fluctuation analysis, we investigate the statistic
properties of traffic rate process (i.e. depicted by the number of
packets per time unit) based on this traffic model, which is in
good accordance with the empirical data.

\section{Model}
In this paper, we treat all the nodes as both hosts and routers
\cite{Guimera,Zhou2006}. The model is described as follows: at
each time step, there are $R$ packets generated in the system,
with randomly chosen sources and destinations. It is assumed that
all the routers have the same capabilities in delivering and
handling information packets, that is, at each time step all the
nodes can deliver at most $C$ packets one step toward their
destinations according to the routing strategy. We set $C=1$ for
simplicity. A packet, once reaching its destination, is removed
from the system. We are most interested in the critical value
$R_c$ where a phase transition takes place from free flow to
congested traffic. This critical value can best reflect the
maximum capability of a system handling its traffic. In
particular, for $R<R_c$, the numbers of created and delivered
packets are balanced, leading to a steady free traffic flow. For
$R>R_c$, traffic congestion occurs as the number of accumulated
packets increases with time, simply for that the capacities of
nodes for delivering packets are limited. To characterize the
phase transition, we use the following order parameter
\begin{equation}
H(R)=\lim_{t \rightarrow \infty} \frac{C}{R} \frac{\langle \Delta
W \rangle}{\Delta t},
\end{equation}
where $\Delta W = W(t+\Delta t) - W(t)$, with $\langle \cdots
\rangle$ indicating average over time windows of width $\Delta t$,
and $W(t)$ is the total number of packets in the network at time
$t$. Clearly, $H$ equals zero in the free-flow state, and becomes
positive when $R$ exceeds $R_c$.

\begin{figure}
\begin{center}
\scalebox{0.8}[0.8]{\includegraphics{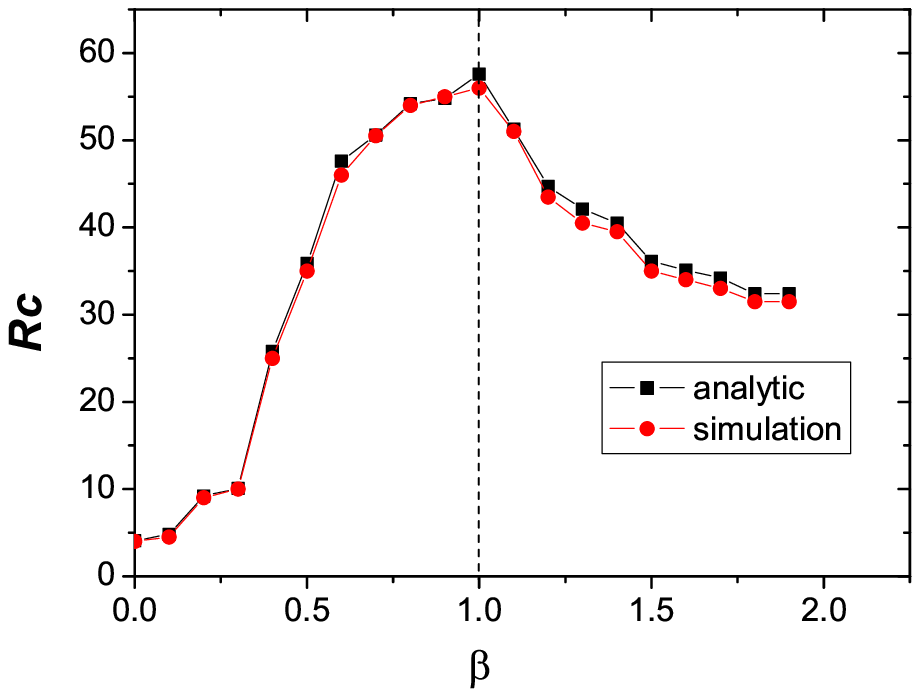}}\\
\scalebox{0.8}[0.8]{\includegraphics{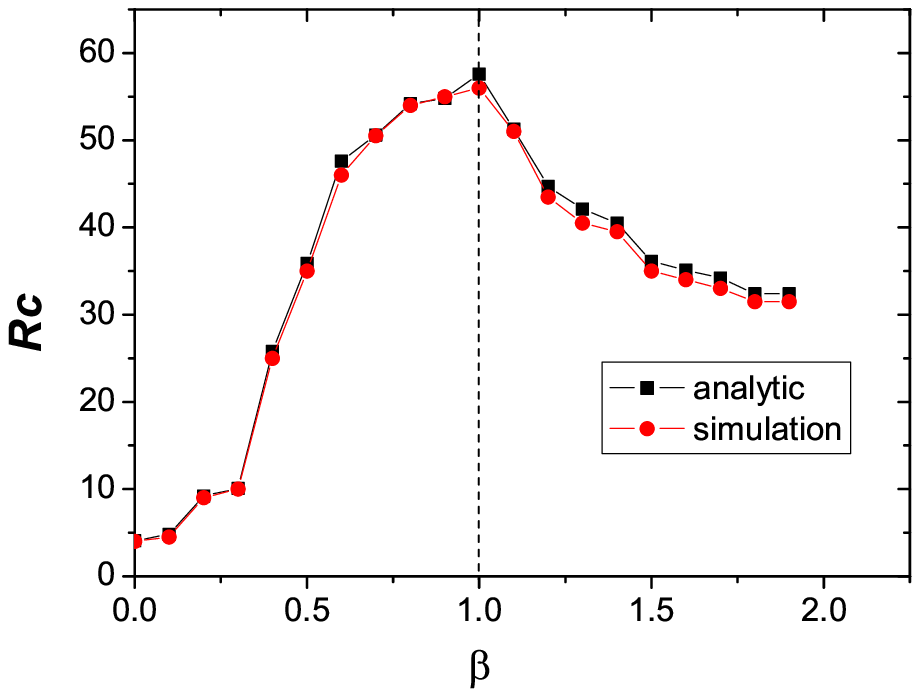}} \caption{(Color online)
The critical $R_{c}$ vs $\beta$ for BA network with size $N=1225$
(up panel) and $N=1500$ (down panel). Both the simulation (black
squares) and analysis (red circles) demonstrate that the maximal
$R_{c}$ corresponds to $\beta \approx 1.0$. The results are the
average over 10 independent runs.}
\end{center}
\end{figure}

\begin{figure}
\begin{center}
\scalebox{0.8}[0.8]{\includegraphics{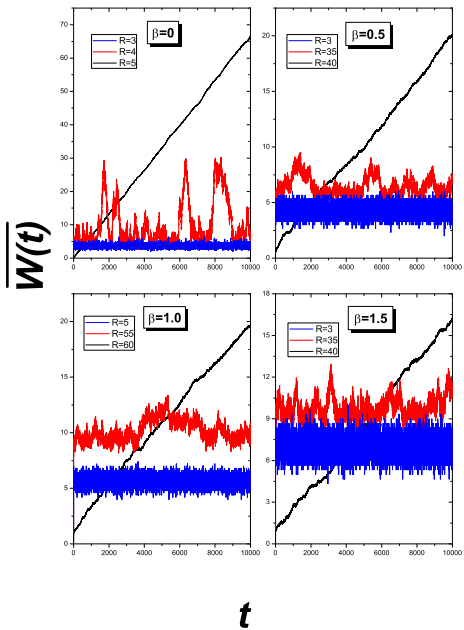}} \caption{(Color online)
The traffic rate process for free (red), critical (blue) and
congested (black) states with different $\beta$.}
\end{center}
\end{figure}

\begin{figure}
\begin{center}
\scalebox{0.8}[0.8]{\includegraphics{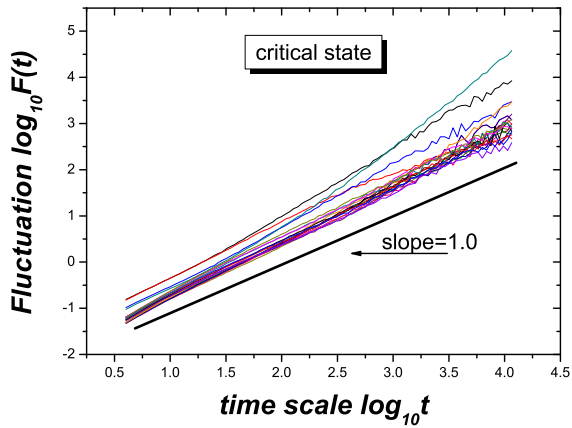}} \caption{(Color online)
The detrended fluctuation analysis of the traffic rate processes
generated by the present model. All the data are obtained from the
critical state, and the different curves represent the cases of
different $\beta$ from 0 to 1.9 at step 0.1.}
\end{center}
\end{figure}

Consider the routing under global protocol \cite{Yan2006}, where
the whole structural information is available, and the fixed
routing algorithm is the most widely used one for its obvious
advantages in economical and technical costs
\cite{Tanenbaum,Huitema}. Actually, the path with shortest length
is not necessarily the quickest way, considering the presence of
possible traffic congestion and waiting time along the shortest
path. Obviously, nodes with larger degree are more likely to bear
traffic congestion, thus a packet will by average spend more
waiting time to pass through a high-degree node. All too often,
bypassing those high-degree nodes, a packet may reach its
destination quicker than taking the shortest path. In order to
find the optimal routing strategy, we define the \emph{efficient
path} \cite{Yan2006}. For any path between nodes $i$ and $j$ as
$P(i\rightarrow j):=i\equiv x_0, x_1, \cdots x_{n-1}, x_n \equiv
j$, denote
\begin{equation}
L(P(i\rightarrow j):\beta)=\sum_{i=0}^{n-1}k(x_i)^{\beta},
\end{equation}
where $k(x_i)$ denotes the degree of the node $x_i$, and $\beta$
is a tunable parameter. The efficient path between $i$ and $j$ is
corresponding to the route that makes the sum $L(P(i\rightarrow
j):\beta)$ minimum. Obviously, $L_{min}(\beta=0)$ recovers the
traditionally shortest path length. As for any pair of source and
destination, there may be several efficient paths between them. We
randomly choose one of them and put it into the fixed routing
table which is followed by all the information packets.

\section{Simulation and analysis}
Some simulation and analytic results are presented in this
section. Without particular statement, all the simulations are
based on the Barab\'asi-Albert (BA) networks \cite{Barabasi} with
average degree $\langle k \rangle=4$.

\subsection{Phase transition}
The packets handling and delivering capacity of the whole network
is an index of performance in traffic system, which can be
measured by the critical value $R_{c}$. At the critical value
$R_{c}$, a continuous transition will be observation from the
free-flow phase to the congested phase. Fig. 1 shows the order
parameter $H$ versus $R$ with different $\beta$, in which the
obvious phase transitions occur. For different $\beta$, it is easy
to find that the capacities of systems are much different.

\subsection{Optimal routing strategy}
We select packet routing strategy through a tunable parameter
$\beta$, and the capacity of whole network are much different. A
natural question arises: Which value of $\beta$ will lead to the
maximal capacity of the network. In Fig. 2, we report the
simulation results for the critical value $R_{c}$ as a function of
$\beta$ on BA networks with the size $N=1225$ and $N=1500$, which
demonstrate that the optimal routing strategy is corresponding to
$\beta=1$ and the size of BA network doesn't affect the optimal
value. In comparison with the traditional routing strategy (i.e.
$\beta=0$), the capacity $R_c$ of the whole network is greatly
improved more than 10 times without any increase in algorithmic
complexity.

\begin{figure}
\begin{center}
\scalebox{0.8}[0.8]{\includegraphics{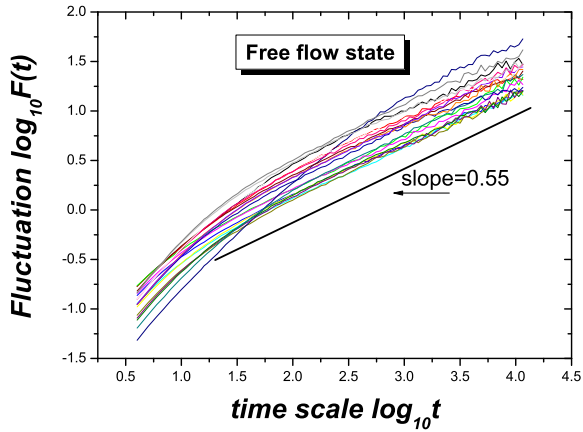}} \caption{(Color online)
The detrended fluctuation analysis of the traffic rate processes
generated by the present model. All the data are obtained from the
free-flow state, and the different curves represent the cases of
different $\beta$ from 0 to 1.9 at step 0.1.}
\end{center}
\end{figure}

By extending the concept of \emph{betweenness centrality}
\cite{Newman2001,Note} to \emph{efficient betweenness centrality}
\cite{ex}, the analytic results can be obtained according to the
Little's law (One can find the concept of efficient betweenness
and the details about the analytical approach in Refs.
\cite{Guimera,Zhao2005,Yan2006}). The analytical results are also
shown in Fig. 2, which agree very well with the simulations.

\section{Detrended fluctuation analysis on the scaling behavior of the traffic rate
process}

\subsection{The critical phenomena of the traffic rate process}

The empirical studies on communication networks confirm the
existence of critical phenomena in traffic rate process
\cite{Takayasu1999,Takayasu2000}. The Congestion duration length
distribution obeys a power-law form with an approximate exponent
-1 in critical state, which implies the self-similarity scaling
exponent (Hurst exponent) $H\approx 1$. Actually, $H<1/2$ stands
for an anti-persistent long-range correlated process in which a
step in one direction is preferentially followed by a reversal of
direction, while $H>1/2$ is interpreted as a persistent long-range
correlated process in which a step in one direction is
preferentially followed by another step in the same direction. A
value of $H=1/2$ is interpreted as the ordinary diffusion (random
walk), in which each step is independent of its preceding one.

In Fig. 3, we report the average number of packets over all the
nodes, $\overline{W}(t)=W(t)/N$, as a time series in free,
critical and congested states, respectively. As time goes on,
$\overline{W}(t)$ in the congested state will become too large to
be plotted together with those in free and critical state.
Therefore, each curve representing the congested state in Fig. 3
is anteriorly divided by a very large number. Those numbers are
not the same for different simulations, and do not have any
physical meanings, they are just used to make the figure more
clear. The behaviors of $\overline{W}(t)$ in the free and
congested states are very simple: In the former case, it
fluctuates slightly around a very low value, while in the latter
case, it increases linearly. However, the time series at the
critical point is very complicated, we next detect its scaling
property by using the detrended fluctuation analysis.


\subsection{Detrended fluctuation analysis}
The empirical study on the self-similar scaling behavior of
traffic rate process can be firstly found in pioneer references
\cite{Leland1993,Erramilli}. The autocorrelation function and
power spectrum are widely used to analyse the the self-similar
scaling behavior of data \cite{Park,Holt,Yao,Karagiannis}.
However, it is shown that all the above methods don't work very
well for the effect of non-stationary, and are less accurate than
the detrended fluctuation analysis (DFA)
\cite{Peng1994,Peng1995,Liu1999}, which has now been accepted as
an important time series analysis approach and widely used
especially for financial and biological data
\cite{DFA1,DFA2,DFA3,DFA4,DFA5}.

The DFA method is based on the idea that a correlated time series
can be mapped to a self-similar process by an integration.
Therefore, measuring the self-similar feature can indirectly tell
us the information about the correlation properties
\cite{Peng1994,Peng1995}. Briefly, the description of the DFA
algorithm involves the following steps.

(1) Consider a time series $x_{i}, i=1, \cdots, N$, where $N$ is
the length of this series. Determine the \emph{profile}
\begin{equation}
y(i)=\sum_{k=1}^{i}[x_{k}-\langle x\rangle], i=1, \cdots, N,
\end{equation}
where
\begin{equation}
\langle x\rangle=\frac{1}{N}\sum_{i=1}^{N}x_{i}.
\end{equation}

(2) Divide profile $y(i)$ into non-overlapping boxes of equal size
$t$.

(3) Calculate the local trend $y_{\texttt{fit}}$ in each box of
size $t$ by a least-square fit of the series, and the detrended
fluctuation function is given as
\begin{equation}
Y(k)= y(k)-y_{\texttt{fit}}(k).
\end{equation}

(4) For a given box size $t$, we calculate the root-mean-square
fluctuation
\begin{equation}
F(t)=\sqrt{\frac{1}{N}\sum_{1}^{N}[Y(k)]^{2}},
\end{equation}
and repeat the above computation for different box sizes $t$ (i.e.
different scales) to provide a relationship between $F$ and $t$.
If the curve $F(t)$ in a log-log plot displays a straight line,
then $F(t)$ obeys the power-law form $t^{H}$ with $H$ the slope of
this line.

As shown in Fig. 4, the mimic traffic rate process also exhibits
the self-similar scaling behaviors at the critical point of phase
transition. The scaling exponents calculated with DFA for
different $\beta$ are approximate $H\approx 1$, and the value of
$\beta$ has almost no effect on $H$. This value of $H$ implies the
$1/f$-type scaling in the power spectrum and the long-range
correlated behavior in a wide range of scales. A very recent
empirical study on the traffic rate process of a University
Ethernet has demonstrated that the real Ethernet traffic displays
a self-similarity behavior with scaling exponent $\approx 0.98$
\cite{ZhouPL2006}, which agrees well with the present result
$H\approx 1$.

In order to confirm that the critical behavior only exists in the
critical state (as is shown in some previuos empirical studies)
\cite{Takayasu1996,Fukuda1999,Takayasu1999,Takayasu2000}, we
analysis the traffic rate processes in the free-flow state. Fig. 5
suggests that traffic rate processes in free state are completely
different from those in the critical state, which exhibit a very
weak long-range correlated behavior with exponent $H=0.55$.
Clearly, if the transporting time of each packet is exactly the
same as the path length (i.e. a packet does not need to wait at
any nodes in the path from its source to destination), there
should be no long-range correlation and the Hurst exponent will be
$H=0.5$. In the present model, even in the free-flow state, a
packet may wait some time steps at some nodes before reaching to
its destination. That is the reason leading to the weak long-range
correlation. However, the waiting effect in the free-flow state is
trifling for the Hurst exponent is close to 0.5.

\section{conclusion}
In conclusion, we have proposed a traffic model based on packet
routing strategy aiming at efficient paths instead of the shortest
paths. This work may be useful for designing communication
protocols for highly heterogeneous networks since the present
strategy can sharply enhance the throughput without any increase
in its algorithmic complexity. The traffic rate process generated
by this model exhibits critical self-similar scaling behavior with
exponent $\approx 1$, which implies the $1/f$-type scaling in the
power spectrum and the long-range correlated behavior in a wide
range of scales. The scaling behaviors of the present model is in
good accordance with the empirical data, thus this model may have
got hold of some key ingredients of the underlying mechanism of
real traffic.

\begin{acknowledgements}
This work was partially supported by the National Natural Science
Foundation of China under Grant Nos. 70471033, 10472116, 10532060,
70571074 and 10547004, the Special Research Founds for Theoretical
Physics Frontier Problems under Grant No. A0524701, and the
Specialized Program under the Presidential Funds of the Chinese
Academy of Science.
\end{acknowledgements}


\begin{thebibliography}{Watts}
\bibitem{Watts} D. J. Watts and S. H. Strogatz, Nature (London) \textbf{393}, 440
(1998).
\bibitem{Barabasi} A.-L. Barab\'{a}si and R. Albert, Science \textbf{286}, 509 (1999).
\bibitem{Li} H. Li and M. Maresca, IEEE Trans. Comput. \textbf{38}, 1345 (1989).
\bibitem{Leland} W. E. Leland, M. S. Taqqu, W. Willinger, and D. V. Wilson,
Comput. Commun. Rev. \textbf{23}, 283 (1993).
\bibitem{Taqqu} M. S. Taqqu, W. Willinger, and R. Sherman, Comput. Commun.
Rev. \textbf{27}, 5 (1997).
\bibitem{Crovella} M. E. Crovella and A. Bestavros, IEEE/ACM Trans. Netw. \textbf{5},
835 (1997).
\bibitem{Pastor} R. Pastor-Satorras, A. V\'{a}zquez, and A. Vespignani, Phys. Rev.
Lett. \textbf{87}, 258701 (2001).
\bibitem{Albert1999} R. Albert, H. Jeong, and A.-L. Barab\'{a}si, Nature (London) \textbf{401}, 103 (1999).
\bibitem{Tadic2006} B. Tadi\'{c}, Int. J. Bifurca. \& Chaos (to be published).
\bibitem{Guimera} R. Guimer\'{a}, A. D\'{i}az-Guilera, F. Vega-Redondo, A. Cabrales,
and A. Arenas, Phys. Rev. Lett. \textbf{89}, 248701 (2002).
\bibitem{Zhao2005} L. Zhao, Y.-C. Lai, K. Park, and N. Ye, Phys. Rev. E \textbf{71},
026125 (2005).
\bibitem{Holme2003} P. Holme, Adv. Complex Syst. \textbf{6}, 163 (2003).
\bibitem{Tadic2001} B. Tadi\'{c}, Physica A \textbf{293}, 273 (2001).
\bibitem{Tadic2002} B. Tadi\'{c}, and G. J. Rodgers, Adv. Complex Syst. \textbf{5}, 445
(2002).
\bibitem{Tadic2004} B. Tadi\'{c}, S. Thurner, and G. J. Rodgers, Phys. Rev. E \textbf{69},
036102 (2004).
\bibitem{Yin2006} C.-Y. Yin, B.-H. Wang, W.-X. Wang, T. Zhou, and H.-J. Yang,
Phys. Lett. A \textbf{351}, 220 (2006).
\bibitem{Wang2006} W. -X. Wang, B. -H. Wang, C.-Y. Yin, Y. -B. Xie, and T. Zhou, Phys. Rev. E \textbf{73}, 026111 (2006).
\bibitem{Echenique} P. Echenique, J. G\'{o}ez-Garde$\tilde{a}$s, and Y. Moreno, Phys. Rev.
E \textbf{70}, 056105 (2004).
\bibitem{Braunstein} L. A. Braunstein, S. V. Buldyrev, R. Cohen, S. Havlin, and H.
E. Stanley, Phys. Rev. Lett. \textbf{91}, 168701 (2003).
\bibitem{Leland1994} W. E. Leland, M. S. Taqqu, W. Willinger, and
D . V. Wilson, IEEE/ACM Trans. Netw. \textbf{2}, 1 (1994).
\bibitem{Paxson1997} V. Paxson and S. Floyd, IEEE/ACM Trans. Netw. \textbf{5},
226 (1997).
\bibitem{Feldmann} A. Feldmann, A. C. Gilbert, P. HUang, and W.
Willinger, Comput. Commun. Rev. \textbf{29}, 301 (1999).
\bibitem{Yang2006} C. -X. Yang, S. -M. Jiang, T. Zhou, B. -H. Wang, and P. -L. Zhou, \emph{2006 International Conference on Communications, Circuits and Systems Proceedings}, (IEEE Press, pp.1740-1743, 2006).
\bibitem{Faloutsos} M. Faloutsos, P. Faloutsos, and C. Faloutsos, Comput. Commun. Rev. \textbf{29}, 251 (1999).
\bibitem{Percacci} R. Percacci and A. Vespignani, Eur. Phys. J. B
\textbf{32}, 411 (2003).
\bibitem{Takayasu1996} M. Takayasu, H. Takayasu, T. Sato, Physica A \textbf{233}, 924 (1996).
\bibitem{Fukuda1999} K. Fukuda, H. Takayasu, M. Takayasu, Fractals \textbf{7}, 23 (1999).
\bibitem{Takayasu1999} M. Takayasu, K. Fukuda, H. Takayasu, Physica A \textbf{274}, 140 (1999).
\bibitem{Takayasu2000} M. Takayasu, H. Takayasu, K. Fukuda, Physica A \textbf{277}, 248 (2000).
\bibitem{Zhou2006} T. Zhou, G. Yan, B.-H. Wang, Z.-Q. Fu, B. Hu, C.-P. Zhu, and
W.-X. Wang, Dyn. Contin. Discret. Impuls. Syst. Ser. B-Appl.
Algorithm \textbf{13}, 463 (2006).
\bibitem{Yan2006} G. Yan, T. Zhou, B. Hu,
Z. -Q. Fu, and B. -H. Wang Wang, Phys. Rev. E \textbf{73}, 046108
(2006)
\bibitem{Tanenbaum} A. S. Tanenbaum, \emph{Computer Networks} (Prentice Hall, Engle-wood Cliffs, NJ, 1996).
\bibitem{Huitema} C. Huitema, \emph{Routing in the Internet} (Prentice Hall, Upper
Saddle River, NJ, 2000).
\bibitem{Newman2001} M. E. J. Newman, Phys. Rev. E \textbf{64}, 016132 (2001).
\bibitem{Note} T. Zhou, J. -G. Liu, and B. -H. Wang, Chin. Phys.
Lett. \textbf{23}, 2327 (2006).
\bibitem{ex} Using the efficient paths instead of the shortest paths in the definition of betweenness
centrality.
\bibitem{Leland1993} W. E. Leland, M. S. Taqq, W. Willinger, and D. V. Wilson, Proc. ACM SIGCOMM, San
Francisco, CA, USA, \textbf{23}(4), 183 (1993).
\bibitem{Erramilli} A. Erramilli, O. Narayan, and W. Willinger, IEEE/ACM Trans
Netw. \textbf{4}, 209 (1996).
\bibitem{Park} K. Park and W. Willinger, \emph{Self-Similar Network Traffic: An
Overview}, in \emph{Self-Similar Network Traffic and Performance
Evaluation}, (Wiley-Interscience, pp.1-39, 2000).
\bibitem{Holt} A. Holt, IEE Proc.-Commun. \textbf{147}(6), 317 (2000).
\bibitem{Yao} L. Yao, M. Agapie, J. Ganbar and M. Doroslovacki, Communications, ICC '03 IEEE International Conference,
pp.1611 (2003).
\bibitem{Karagiannis} T. Karagiannis, M. Molle and M. Faloutsos, IEEE Internet Computing \textbf{8}(5), 57 (2004).
\bibitem{Peng1994} C. K. Peng, S. V. Buldyrev, S. Havlin, M. Simons, H. E.
Stanley and A. L. Goldberger, Phys. Rev. E, \textbf{49}, 1685
(1994).
\bibitem{Peng1995} C. K. Peng, S. Havlin, H. E. Stanley and A. L.
Goldberger, Chaos \textbf{5}, 82 (1995).
\bibitem{Liu1999} Y. H. Liu, P. Gopikrishnan, P. Cizeau, M. Meyer, C. K. Peng and H. E.
Stanley, Phys. Rev. E, \textbf{49}, 1390 (1999).
\bibitem{DFA1} A. Bunde, S. Havlin, J. W. Kantelhardt, T. Penzel,
J. -H. Peter, and K. Voigt, Phys. Rev. Lett., \textbf{85}, 3736
(2000).
\bibitem{DFA2} L. M. Xu, P. C. Ivanov, K. Hu, Z. Chen, A. Carbone,
and H. E. Stanley, Phys. Rev. E, \textbf{71}, 051101 (2005).
\bibitem{DFA3} Z. Chen, K. Hu, P. Carpena, P. Bernaola-Galvan, H.
E. Stanley, and P. C. Ivanov, Phys. Rev. E, \textbf{71}, 011104
(2005).
\bibitem{DFA4} H. -J. Yang, F. -C. Zhao, L. -Y. Qi, and B. -L. Hu,
Phys. Rev. E, \textbf{69}, 066104 (2004).
\bibitem{DFA5} S. -M. Cai, P. -L. Zhou, H. -J. Yang, C. -X. Yang,
B. -H. Wang, and T. Zhou, Chin. Phys. Lett. \textbf{23}, 754
(2006).
\bibitem{ZhouPL2006} P. -L. Zhou, S. -M. Cai, T. Zhou, and Z. -Q. Fu, \emph{2006 International Conference on Communications, Circuits and Systems Proceedings}, (IEEE Press, pp.1744-1748, 2006).
\end{thebibliography}
\end{document}